\documentclass[12pt]{article}
\usepackage{graphicx}

\begin{document}

\title{Modeling controversies in the press:\\
the case of the abnormal bees'death}
\author{Alexandre Delano\"e\thanks{alexandre.delanoe@telecom-paristech.fr},\\
LTCI, \'Ecole T\'el\'ecom ParisTech, \\Rue Barrault, 46, 75013 Paris, France\\Serge Galam\thanks{serge.galam@polytechnique.edu},\\
Centre de Recherche en \'Epist\'emologie Appliqu\'ee,\\
\'Ecole Polytechnique and CNRS, France
}

\date{}
\maketitle

\begin{abstract}

The controversy about the cause(s) of abnormal death of bee colonies
in France is investigated through an extensive analysis of the french
speaking press. A statistical analysis of textual data is first
performed on the lexicon used by journalists to describe the facts
and to present associated informations during the period 1998-2010.
Three states are identified to explain the phenomenon. The first state
asserts a unique cause, the second one focuses on multifactor causes
and the third one states the absence of current proof. Assigning
each article to one of the three states, we are able to follow the
associated opinion dynamics among the journalists over 13 years. Then,
we apply the Galam sequential probabilistic model of opinion dynamic
to those data. Assuming journalists are either open mind or inflexible
about their respective opinions, the results are reproduced precisely
provided we account for a series of annual changes in the proportions
of respective inflexibles. The results shed a new counter intuitive
light on the various pressure supposed to apply on the journalists by
either chemical industries or beekeepers and experts or politicians. The
obtained dynamics of respective inflexibles shows the possible effect of
lobbying, the inertia of the debate and the net advantage gained by the
first whistleblowers.

\end{abstract}

\newpage

\section{Introduction}

Public opinion is a key feature to determine which decisions should
be taken by policy makers in modern democratic societies especially
on subjects dealing with public and environmental risks. Therefore
the understanding of its underlining mechanisms becomes at the top
priority among current major challenges. In that context, the study
case of the precautionary principle is a sensitive issue as it states
that if a policy has a suspected risk of causing harm to the public
or to the environment, in the absence of scientific consensus, then
the action or policy is harmful. But, since debates are driven by
incomplete scientific data, the question becomes: how "suspected" risks
are translated \cite{Callon:1986} in the public debate?

Such issue of opinion dynamics has already been evaluated from a
theoretical viewpoint, as a diffusing dynamics which leads to the success
-- or the disappearance -- of one of the opinions in competition.
In such case, the dynamics is understood as as selection process.
But yet there remains the empirical challenge which aims to compare
data descriptive analysis with theoretical modeling. Following this
objective, this article first presents the context of the controversy
that feeds a text-mining analysis on a press corpus in order to
introduce a simulation framework dealing with opinion dynamics.

\section{Setting the problem}

The abnormal bees' death\footnote{Called colony collapse disorder (CCD)
in others country (like in US).} around the world appears as a complex
issue that has been especially controversial in French daily press since
1998. This fact can be explained by many factors as it has been showed
by scientific studies till 2012. Because of this complexity, the focus
is made on the French public debate which has been analyzed with a mixed
methodology.

After collecting all papers dealing with the ecological concern, a
statistical analysis of textual data has been performed on the lexicon
used by the authors to describe and report associated information. This
step of the research aims to throw light on 3 points:

\begin{enumerate}
\item Define (in time and space) the particular social phenomenon;
\item Show the salient paradoxical dynamics that has to be reproduced with the model;
\item Select the relevant (and simplified) entities (especially between journals and
authors level) that justify the use of a model.
\end{enumerate}

The data are then reproduced with the Galam sequential probabilistic
model of opinion dynamics. The proportions of inflexibles agents are
finally confronted with the results of the text mining analysis.

\subsection{About the social phenomenon}

Two heuristics, depending on the definition of the issue, could guide
our investigation. Only one way will be chosen for the modeling. 

Bees' death may appear as a tangible proof~\cite{Chateauraynaud:2004}
of the \textit{uncontrolled} pesticides' toxicity. In such heuristic
axis, the focus is especially made on the pesticides' case usually
called "Gaucho" affair \cite{Maxim:Sluijs:2010} " or "no Gaucho nor
Regent \cite{Delanoe:2004} \cite{Delanoe:2007}" since Gaucho and Regent
are the names of the chemicals accused by stakeholders from 1998 to 2004
years. Sociology of the controversy focused on pesticides in such way.

But the issue may have a different formulation for a new heuristic
approach. The focus has not to be made only on the pesticides themselves
but especially on the bees. The problem formulation becomes: "how to
explain the death of the bees?" We follow this acceptation since such
description throws light on the consequences of abnormality that define
the public problem~\cite{dewey:1927} (which evolves from a personal
viewpoint to a public stake since, according to the whistleblowers,
the bees' death involves all publics). In such issue, the causality of
the death's bees is investigated by the actors (beekeepers in first)
in order to unveil responsibilities~\cite{Gusfield:1981:eng} for this
hazard that becomes collective.

\begin{figure}[h]
\begin{center}
\includegraphics*[width=12cm,height=9cm]{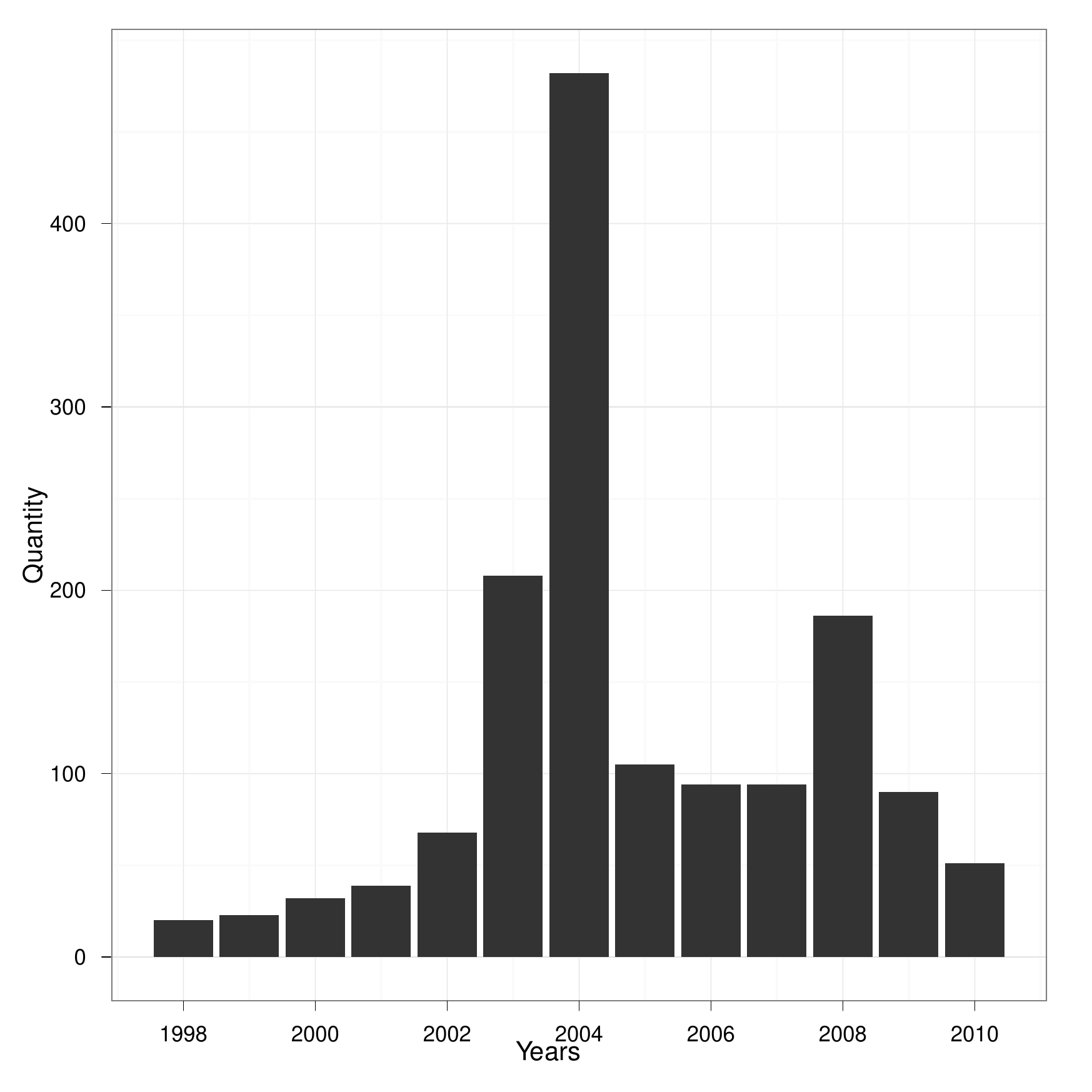}
\caption{Quantity of articles published each year by French daily,
weekly and monthly press dealing with the bees' death (N=1467).}
\label{history}
\end{center}
\end{figure}

To study this \textit{public problem}, text-mining with descriptive
statistics describe quantitatively the main textual features of our
collection of data which is based on daily, weekly and monthly press.
This corpus brings together all articles that deal with the death of
the bees from 1998 to 2010 years in French language. Articles published
during this period have been extracted from complementary databases --
Lexis-Nexis and Factiva -- in order to cover the event the best as we
can. As a result, 1467 articles have been collected for the text mining
analysis (Fig.~\ref{history}).

\section{The salient paradoxical dynamic}

The first step on the study shows how \textit{judgement dynamics} are
connected to the claim of unusual death of bees through the evolution of
explanatory factors reported in the media. Two periods are distinguished
when scientific knowledge is translated in public debate. Firstly agents
focus on unifactor cause. Secondly, agents focus on multifactor causes.

Indeed, we have first formulated the hypothesis that modeling
the opinion dynamics can be tackled through the prism of factors
written by journalists to report the controversy and to explain
the death of the bees. During the public debate many factors have
been identified~\cite{afssa:2008}, that is why all papers have been
classified in 3 categories according to the factors reported in papers:

\begin{enumerate}
\item articles which mention only a unifactor explanation (caused by pesticides);
\item articles that assert there is no proof yet (that pesticides cause damages to bees);
\item articles that highlight on multifactor explanation of the phenomena (pesticides + one other factor at less).
\end{enumerate}

\begin{figure}[h]
\begin{center}
\includegraphics*[width=12cm,height=9cm]{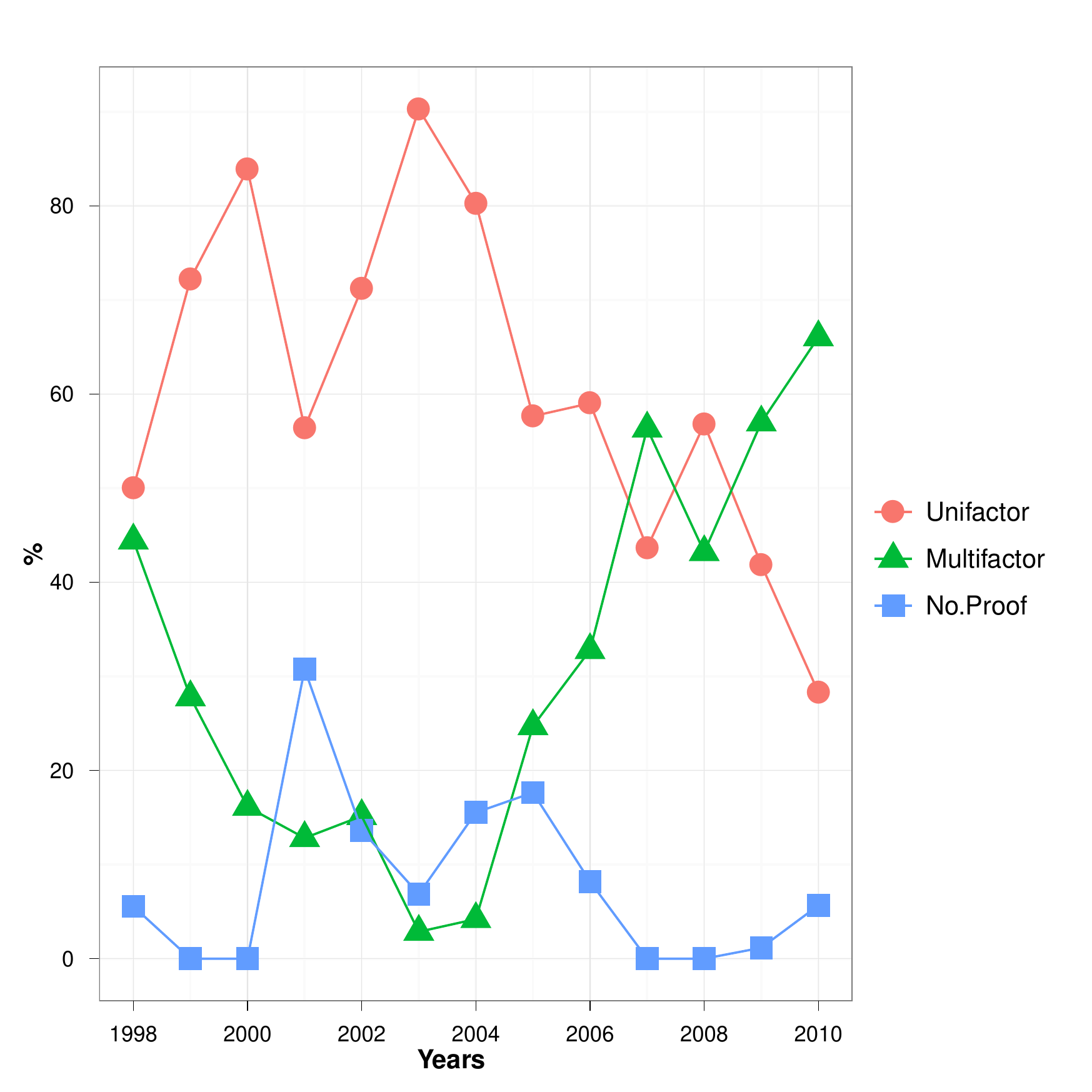}
\caption{Proportion of articles published each year by daily press
dealing with unifactor, multifactor or "no proof" categories. 85.3\% of
the corpus (i.e. 1252 articles) has been tagged in that way.}
\label{factors}
\end{center}
\end{figure}

85.3\% articles of the corpus have been tagged with semi-automatic
textual analysis tools and with human reading for validation.
The chronology of these categories can be summarized by the
Histogram~\ref{factors}.

The Figure~\ref{factors} shows the proportions of categorized papers
published per year. The causality reported in the articles is balanced
between unifactor and multifactor explanations at time T1 (1998
year). Next, the unifactor thesis has been spreading at time T2
(1999-2000) whereas the no-proof assertion has been asserted at time
T3 (2001). Then the unifactor cause has been spreading again at time T4
(2002-2004 years). After the event of precaution principle application
on pesticides in 2004, a multifactor explanation has been increasing
at time T5 (2005-2007 years). At time T6 (2008), the unifactor thesis
has been highlighted again. Finally, at time T7 (2009-2010), the
multi-factors papers have been mainly written. If "controversy" means
a shift in the trend of meanings in the public debate, then this study
consists in explaining such shift. How these proportions may change?

The key dates make it easier the understanding of the
facts\footnote{This chronology explains why the segmentation by years
has been used for the text mining analysis. Indeed this segmentation has
effect on the dynamics detected. The event is long enough to analyze
it by years since each (micro) event is "automatically" indexed inside
one year.}. Actually 2 large trends draw the main dynamics. The period
from 1998 to 2004 represents the first trend of the controversy which
tends to focuses on one cause: the chemicals. The second period, from
the 2005 to 2010 years, deals with multi-factors such as environment,
parasites, viruses or new insecticides.

Since 1998 beekeepers have assumed the role of whistleblowers
focusing on one main factor, the neuro active insecticides called
"neonicotinoids". After controversial discussions on incomplete
scientific data studies and a rough debate during the political
elections, because of or thanks to the precautionary principle, in
2004 chemical products have been banned from sale by the ministry of
agriculture in order to solve this disturbing fact. After this event,
multifactor causes are highlighted in scientific field, and then in the
press, showing a shift in the interpretation of the causes of the death.
After 2005, multifactor causes are mentioned until 2010: the debate does
not focus on the issue of the bees' death any more (and its consequences
for food economy, environment and humanity) but on the premisses of the
reasoning that lead to the multifactor causes.

Before introducing the model, relevant social entities have to be
selected. Even if actors involved in this debate are beekeepers,
politics, experts/scientists and industrialists -- who are selling
chemicals\footnote{Chemicals are named Regent and Gaucho from 1998 to
2004 and after Cruiser from 2005 to 2010.}--, the journalists, as third
party, are the agents writing the papers.

\section{Selecting the relevant entities}

To tackle this public issue in French daily press with the framework
of opinion dynamics, it might be useful to provide a new light on
agents who are implied in this phenomenon. In this case the agents are
exclusively authors since the methodology described above can not take
into account the opinion of the readers or the others agents. Behind the
meta-categories defined previously (unifactor, multifactor, no-proof),
what kind of social entities are implied to explain the dynamics? Focus
is then made on journals and authors level to better describe the
phenomenon.

\subsection{The journal level}

About 59 journals participated to the debate but only a few really
"inflected" the opinion dynamics. Regional papers first published on
unifactor causes and their opinion has been relayed by specific national
papers.

\begin{figure}[h]
\begin{center}
\includegraphics*[width=12cm,height=9cm]{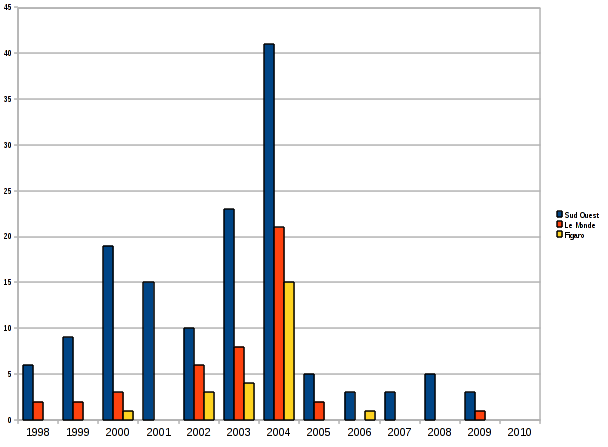}
\caption{Quantity of unifactor papers per journal through time}
\label{unifactor}
\end{center}
\end{figure}

\begin{figure}[h]
\begin{center}
\includegraphics*[width=12cm,height=9cm]{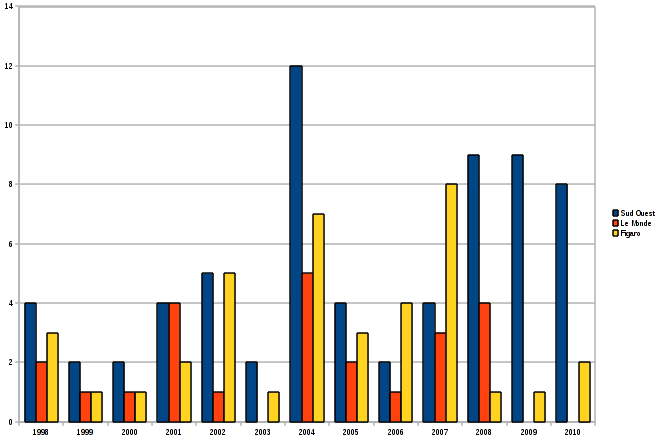}
\caption{Quantity of multifactor papers per journal through time}
\label{multifactor}
\end{center}
\end{figure}

Focus is first made on journals asserting unifactor causes
(Fig.~\ref{unifactor}). The regional journal called "Sud
Ouest~\footnote{Sud Ouest is followed by Ouest-France, which published
unifactor papers in the same proportions.}" is the first journal to
mainly publish articles on unifactor cause. Sud Ouest is quickly
followed by the national journal "Le Monde" during the 2002-2003
years. Sud Ouest appears as an "inflexible" agent who has been mainly
supporting the same thesis along the first period. The journal "Le
Monde" appears more flexible, as "open minded agent", because proportion
of multifactor papers gradually decrease whereas unifactor cause
is gradually reported (from 1998 to 2002). If we focus on journals
assuming multifactor causes (Fig.~\ref{multifactor}), during all the
public debate, Le Figaro is almost the main journal defending the
industrialists since its papers has been supporting the idea that others
factors must be taken into account. But such granularity of social
entity does not detect if there is an editorial opinion or individual
opinion that could explain the dynamics.

\subsection{The individuals or authors'level}

\begin{figure}[h]
\begin{center}
\includegraphics*[width=12cm,height=9cm]{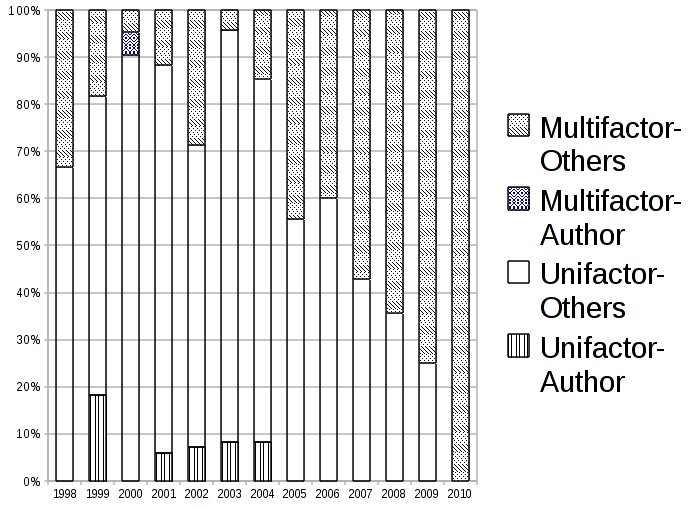}
\caption{Sud Ouest Newspaper, main author and others authors
contribution according to multifactor or unifactor reports, N=118 articles}
\label{sudouest}
\end{center}
\end{figure}

\begin{figure}[h]
\begin{center}
\includegraphics*[width=12cm,height=9cm]{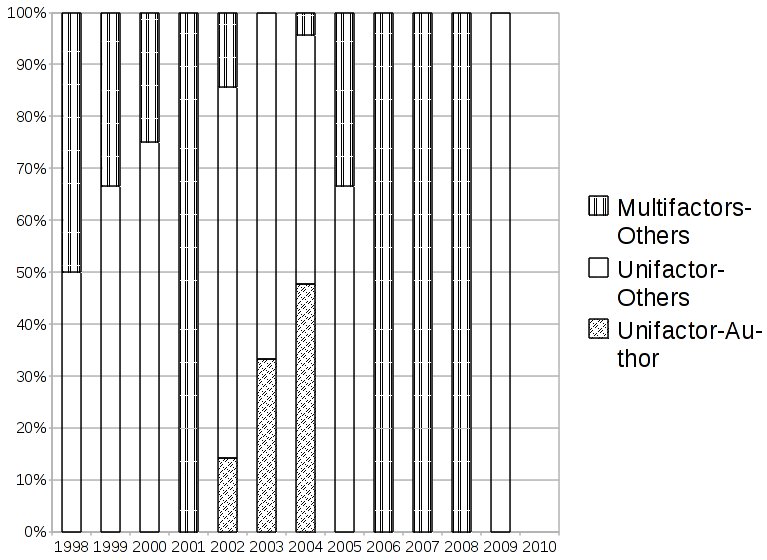}
\caption{Le Monde Newspaper, main author and others authors contribution
according to multifactor or unifactor reports, N=62 articles}
\label{lemonde}
\end{center}
\end{figure}

\begin{figure}[h]
\begin{center}
\includegraphics*[width=12cm,height=9cm]{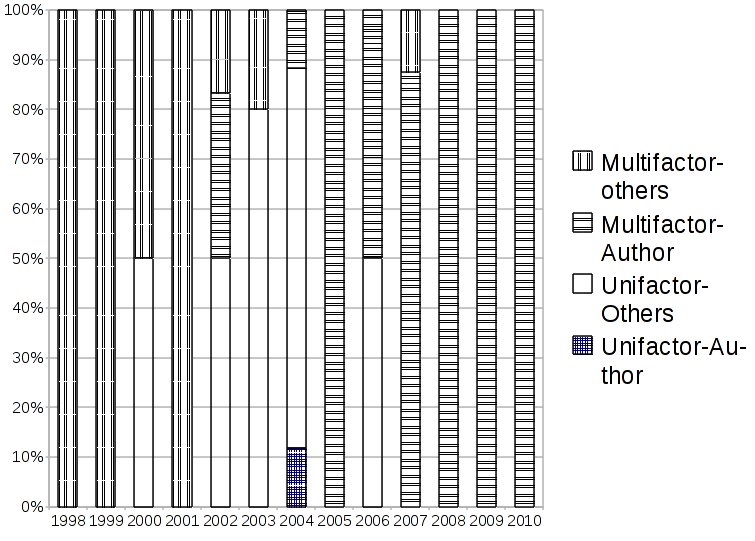}
\caption{Le Figaro Newspaper, main author and others authors contribution
according to multifactor or unifactor reports, N=50 articles}
\label{lefigaro}
\end{center}
\end{figure}

Each newspaper has more than one author writing papers dealing with the
bees' death. Hence, if we consider author's profiles, we have to detect
the main author in order to compare his writings with the papers written
by others journalists~\footnote{To simplify we suppose that author and
journalist is almost the same.}, inside the same newspaper obviously.
We call "main author", the journalist who has published more articles
than all the others individually supposing all the other agents do not
change during the considered period (in such case it would be a change
of population and not a change of the agents themselves). According to
this methodology, we can throw light on 3 profiles which seem to be 3
ideal types~\footnote{According to Max Weber definition.}:

\begin{enumerate}

  \item The main author has constantly published unifactor papers
  in the same way as all others journalists during the same period
  (Fig.~\ref{sudouest}). This is the case of Sud Ouest. Hence an
  inflexible and editorial position can be assumed.

    \item The main author has increasingly published unifactor papers
  during the first period whereas others journalists have published
  unifactor papers during the first period \textit{and} multifactor
  papers during the second period (Fig.~\ref{lemonde}). This is the case
  of Le Monde. Hence the main author appears as an inflexible unifactor
  author whereas the others authors appear more flexible.

  \item The main author has published mainly multifactor papers whereas
  others journalists, in the same newspaper but only during the first
  period, have written unifactor papers (Fig.~\ref{lefigaro}). Hence
  the main author appears as an inflexible multifactor author whereas the
  others authors appear more flexible as an "open minded agents".

\end{enumerate}

According to these results, simulation of the public debate can be
done with inflexible or flexible agents who are composing the authors
of the articles.

\section{Three hypotheses to justify the model}

We do not need to suppose neither meta nor social categories
\textit{a priori} -- such as network, interest, education, skills or
\textit{habitus} -- to explain the dynamics observed since only 3 minimal
hypothesis are needed to justify the model in this paper.

First hypothesis states that agents can distinguish the hierarchy
of factors they have mentioned during the debate. Suppose $F_{1},
... , F_{p}, F_{p+1}, ..., F_{n}$ factors could cause the death of
the bees (in fact more than 40 factors are implied~\cite{afssa:2008}).
In such cognitive rationality~\cite{Boudon:2007} during the first period
only some factors $F_p$ or $F_{p+1}$ or $F_{p+2}$ are involved according
to the agents. Moreover actors infer a hierarchy between them: $ F_p
< F_{p+1} < F_{p+2} $. For example, beekeepers' assertions reported by
the journalists mainly focused on one cause, the chemicals, according
to their own inquiry. If beekeepers focused exclusively on one factor,
the industrialists tried to highlight the fact the level of toxicity
is not high enough to kill the bees; then they have focused on others
factors. At the individual level, each third party -- the journalist in
that case -- have "good reasons enough"~\cite{Boudon:2007} to report the
social fact as an inflexible (i.e. to publish mainly unifactor articles)
or as an more flexible agent with open minded behaviour (changing the
proportion of unifactor and multifactor articles). We do not explain
their reasons but the way how they translate the public debate and the
consequences for the dynamics of the public problem.

To understand the social fact, we also assume a second hypothesis which
supposes that the proportion of inflexible agents in a public debate
may explain the shift in the bees' death explanation along the corpus.
Denoting inflexibles and flexibles agents leads to evoke the Galam
Sequential Probabilistic Model of opinion dynamics \cite{Galam:2002}
\cite{Galam:2005} \cite{Galam:2007} \cite{Galam:2010}. It considers the
social field as a composite of heterogeneous agents with inflexibles and
flexibles agents in order to simplify and mimic the reality. Using a
one-person-one-argument principle, the model implements an opinion shift
via small group discussions monitored by local majority rules. Why does
this model could be adapted to this fact?

A third -- and also minimal -- hypothesis is then supposed:
the cost of defending a thesis may depend on the cost of its
falsification~\cite{Radnitzky:1987:en}. A man of science (and therefore
a journalist in this peculiar case) continues to believe in a theory
if the accumulation of objections makes it too "expansive" the fact
of falsificating it~\cite{Radnitzky:1987:en}. In such epistemological
way, the cost-benefit analysis~\cite{Radnitzky:1986} or in others
words, the fact to defend a thesis depends on his difficulty to
defend it. The problem of this empirical "base" is an investment
problem: whether or not to invest time and effort into processing a
particular basic statement into a falsifying hypothesis for the theory
we wish to test. Then, the valuation of the costs of rejecting or,
as the case may be, of defending a basic statement, are objective.
With respect to this approach, in basic public and scientific issue,
the issue is not one of acceptance or rejection of a single theory,
but rather of theory preference. Both the rational response to a
falsification and rational theory preference may be governed by
cost-benefits-analysis-considerations for the purpose of this model.

\section{Modeling the controversy with the mixed inflexible-flexible agents model}

At this stage to try to understand the origin of the empirical data of
Figure~\ref{factors} one could use a model of opinion dynamics. However
the problem is that most models of opinion dynamics yields threshold
dynamics with attractors which means the dynamics either spread an
opinion or shrink it~\cite{Castellano:2009} \cite{Galam:2008}. Given
some initial conditions with some fixed parameters, the respective
proportions of the various opinions are considered as internal
parameters which evolve till reaching the corresponding attractor.
At odd, Figure~\ref{factors} exhibits a series of brutal reversal of
trends. Focusing on the unifactor curve reversal points appear at
respectively the years 2000, 2001, 2003, 2007, 2008 with a quasi-plateau
between 2005 and 2006.

\begin{figure}[h]
\begin{center}
\includegraphics*[width=12cm,height=9cm]{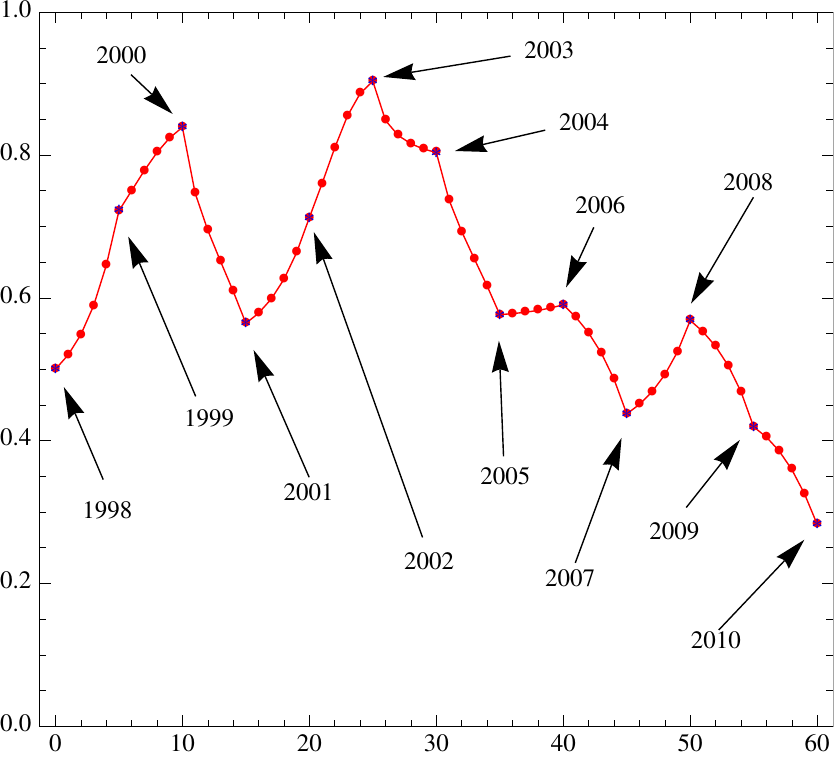}
\caption{Normalization of journals assuming unifactor cause}
\label{fit}
\end{center}
\end{figure}

To implement such breaking in the dynamical trends the values of
the external parameters must be changed. Indeed, we interpret them
as a signature of a redistribution of the respective proportions of
inflexibles in favor of either the unifactor state or the combination of
multifactor and no proof states. Then, using the Galam sequential model
with groups of size 3, we are able to fit the exactly data given some
values of inflexibles. Accordingly, Figure~\ref{proportions} shows the
year evolutions of the various proportions of inflexibles making the
model heuristic to identify among the various journalists the respective
proportions lobbied for a cause.

We start from the data shown in Figure \ref{factors} for the unifactor
proportions. We denote them $P_{A,T}$ with $T=0, 1,..., 12$ for year
1998 $(T=0)$ till 2010 $(T=12)$. The values are $P_{A,0}=0.500,
P_{A,1}=0.722, P_{A,2}=0.839, P_{A,3}=0.564, P_{A,4}=0.712,
P_{A,5}=0.903, P_{A,6}=0.803, P_{A,7}=0.576, P_{A,8}=0.590$,
$P_{A,9}=0.437, P_{A,{10}}=0.568$, $P_{A,{11}}=0.419, P_{A,{12}}=0.283$.

Given some values $p_{A,t}$ and $p_{B,t}=1-p_{A,t}$ at time $t$ the
Galam model of opinion dynamics \cite{Galam:2008} determines the value
$p_{A,t+1}$ and $p_{B,t+1}$ at time $t+1$ associated with one update,
which corresponds to the evolution of individual opinions driven by
local groups discussions during one period of time. From the data it
happens to be reasonable to choose 5 consecutive updates to span a full
year. Such a choice is not unique. However, once the choice is done we
apply five consecutive updates to go from each year to the next one to
span the full period of time from year 1998 till year 2010.

We include the possible existence of inflexibles on both sides whose
proportions are denoted respectively $a$ and $b$. It means that given
a proportion $p_{A,t}$ of supporters of the unifactor cause, only the
fraction $p_{A,t}-a$ is flexible, i.e., able to shift opinion under
convincing arguments with a proportion $a$ of A supporters which never
changes opinion whatever may happen. The same holds for opinion B.

We thus have $p_{A,t}\geq a$ and $1-p_{A,t}\geq b \Longleftrightarrow p_{A,t}\leq 1-b$, which combine to 
\begin{equation}
a\leq p_{A,t} \leq 1-b ,
\label{constraint-ab1}
\end{equation}
to which we add the constraints
\begin{equation}
0\leq a \leq 1  \qquad \mbox{and} \qquad 0\leq b \leq 1 \qquad \mbox{with} \qquad 0\leq a+b  \leq 1 .
\label{constraint-ab2}
\end{equation}
From \cite{Galam:2008} the update equation writes for update groups of size 3,
\begin{equation}
p_{A,t+1}=-2 p_{A,t}^3+(3 +a+b)p_{A,t}^2-2a p_{A,t}^2 +a .
\label{inflex-ab}
\end{equation}

Then, starting from $p_{A,0}=P_{A,0}=0.500$ and iterating Eq.
(\ref{inflex-ab}) five times we want to determine the minimum values
of $(a_0, b_0)$ which yields the best realization of the equality
$p_{A,5}=P_{A,1}$ after five iterations. Indeed we aim at the exact
equality within 3 digits. Next we repeat the scheme finding the values
$(a_1, b_1)$ which leads the equality $p_{A,10}=P_{A,2}$ still within
a 3 digit precision starting from $p_{A,5}$ and so on till $(a_{12},
b_{12})$ to the realization of the equality $p_{A,60}=P_{A,12}$. The
respective values obtained are exhibited in Table \ref{ab}.

\begin{table}[h]
\centering
\caption{inflexible proportions at each year}
\begin{tabular}{|c||c|c||c|c|} 
\hline
 t, T & {$a_t$}  	& $b_t$ & $p_{A,t}$ & $P_{A,t}$ \\
\hline
\hline
0, 1 & 0.080 &	0	&	0.500 &	0.722 \\
\hline
5, 2 & 0	& 	0.117	&	0.722 & 0.839 \\
\hline
10, 3 & 0	& 	0.261	&	0839 & 0.564  \\
\hline
15, 4 & 0	& 	0.055	&	0.564 & 0.712 \\
\hline
20, 5 & 0	& 	0.077	&	0.712 & 0.903 \\
\hline
25, 6 & 0	& 	0.154	&	0.903 & 0.803 \\
\hline
30, 7 & 0	& 	0.251	&	0.803 & 0.576 \\
\hline
35, 8 & 0	& 	0.108	&	0.576 & 0.590 \\
\hline
40, 9 & 0	& 	0.176	&	0.590 & 0.437 \\
\hline
45, 10 & 0.1397	& 	0	&	0.437 & 0.568 \\
\hline
50, 11 & 0    & 	   0.152	&	0.568 & 0.419 \\
\hline
55, 12 & 00746	& 	0	&	0.419 & 0.283 \\
\hline
\end{tabular}
\label{ab}
\end{table}

\begin{figure}[h]
\begin{center}
\includegraphics*[width=12cm,height=9cm]{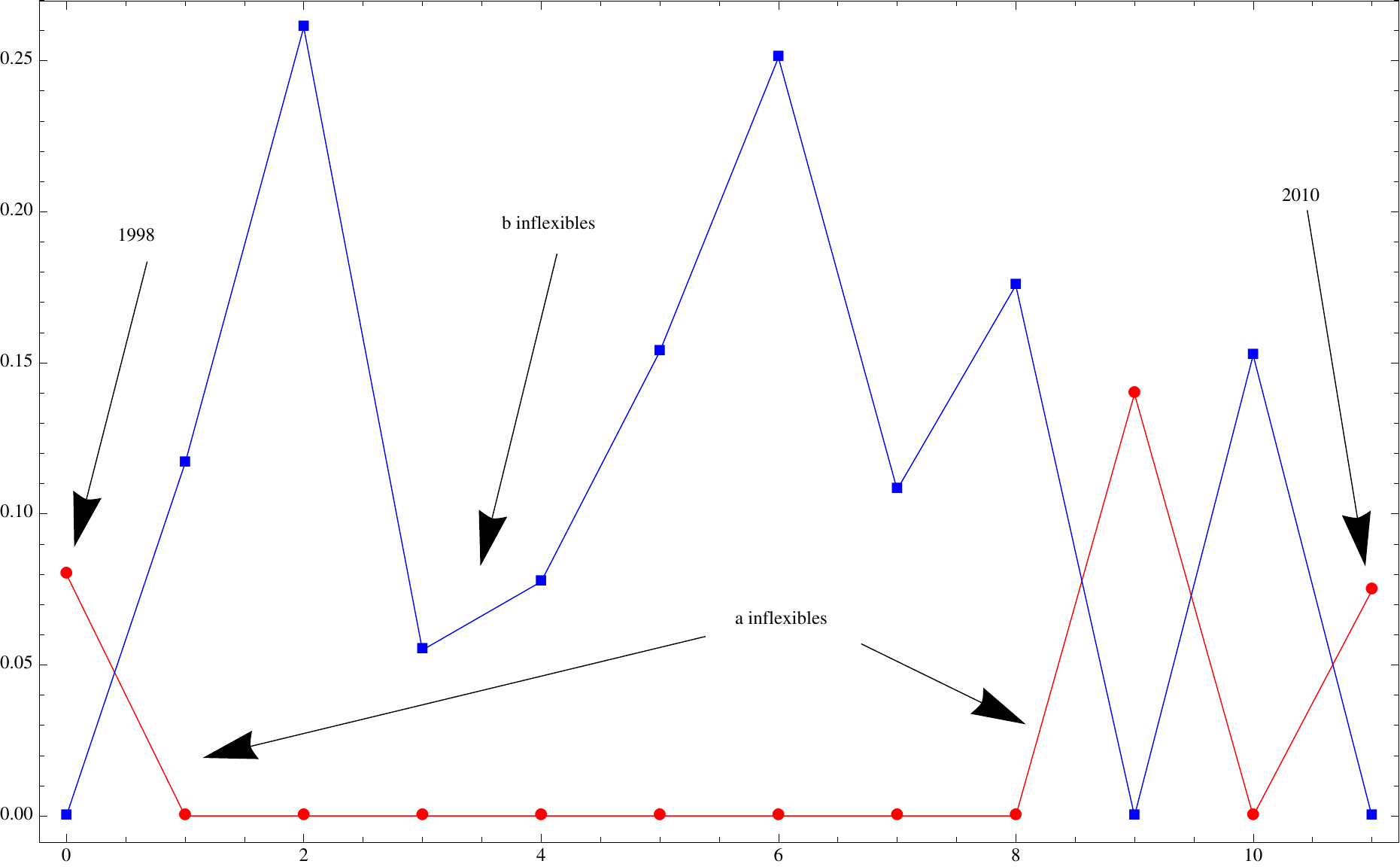}
\caption{Proportions of inflexible agents at each year }
\label{ab+}
\end{center}
\end{figure}

It is worth to stress that once we start from $p_{A,0}=P_{A,0}=0.500$
we repeat Eq. (\ref{inflex-ab}) 60 times in a row changing only the
values $(a, b)$ every five updates. Those values are exhibited in Figure
\ref{ab+}.

The overall picture drawn from our fit sheds an interesting light which
is contrary to what would have been \textit{a priori} expected: the
proportions of multi/uni-factor categorized papers are not the same
as multi/uni-factor-inflexible agents. Most of the "lobbying" effort
has been performed by the multifactor side and not by the unifactor
one. Looking at the data, it is worth to notice that the first point is
located at exactly fifty percent ($p_0=0.50$) which shows a balanced
distribution of the journalist opinions at the starting point of the
controversy.

It seems that the unifactor side was "lucky enough" to have by chance
a few inflexibles on its side (8 $\%$) while none were on the other
side. That has lead the unifactor side to jump to $p_1=0.72$ the next
year. Worried by the situation the other side has been supported by
a few inflexibles on its sides ($12\%$) with a reduction by half of
the unifactor inflexibles ($4\%$). But it was not enough to curb the
spreading of the unifactor side which reached $p_2=0.84$. It is worth
to emphasize that we are using 5 cycles of update per year. At this
desperate level the industrial side made a huge investment in a huge
lobbying getting up to $27\%$ of inflexibles on its side against the
still $4\%$ of unifactor inflexibles. The expected result was indeed
achieved with a falling down to $56\%$ from $88\%$ of the proportion of
articles supporting the unifactor side. Perhaps satisfied with such a
success the industrial side has reduced its lobbying pressure keeping
only $8\%$ of inflexibles against $6\%$ for the unifactor side. However,
not aware of the threshold nature of the dynamics, since $56\%$ is lower
than $50\%$ the next two years brought the unifactor side back to high
values at $p_4=0.71$ and at the top value $p_5=0.90$.

\section{From the model, back to the data}

Each period of the controversy has to be analyzed since each phase of
the debate has its own dynamics which depends also on the terms used by
actors for their argumentation. The first trend (1998-2004) shows a
shift from hypothesis of factors to the certitude of the proof, from
a personal concern to a public hazard, from the interrogation to the
assertion: "the facts have proven that \textit{only} pesticides kill the
bees". The second trend (2005-2010) shows an investigation on the others
causes since one cause has been banned. If the first shift gradually
focuses on the issue of the death's bees, the second one focuses on the
premisses, i.e. all the causes which could explain the death of the
bees.

During the first period 1998-2004, it becomes even harder to falsify
unifactor explanation of abnormal bees' death not only because it is
scientifically right but also because it is right in the morality field.
Indeed, the public debate progressively focused on the consequences of
negative externality of pollution since the precautionary principle has
been reversing the burden of the proof: industrialists must prove the
safety of their products since then can hardly make profit with negative
external effects. In the context of European and Regional political
elections in 2004 year, the death of the bees has been interpreted
as a hazard for the environment and for humanity; using an argument
authority with the name of Einstein who would have said "when bees
disappear, humanity has 4 years left". Evolution of lexical textual
data shows a semantic shift from scientific field to moral accusation,
from analytic verity to moral justice~\cite{Delanoe:2010}. That is why
the proportion of multifactor-inflexible agents increases during the
first phase to counter the framing argumentation firstly launched by the
whistleblowers.

During the second period, 2005-2010, pesticides have been banned
from sale but bees are still dying. New factors have been
investigated~\cite{afssa:2008}. Therefore, it becomes even harder to
falsify multifactor explanation of abnormal bees's death. The number of
unifactor-inflexible agents has been increasing to cope with this trend.
Our corpus arbitrary stops the investigation at the end of 2010 to throw
light on the second trend of the multifactor dynamics but new scientific
studies appeared in 2012 focusing on the pesticides and reviving the
affair.

\section{Conclusion and discussion}

From the above results, this study case shows how empirical data and
theoretical modeling can stimulate each others to tackle the dynamics of
a controversy. four main results can be discussed.

First result has consequences on the methodology to study a controversy.
The particular social phenomenon has been defined with this research
question: how agents do explain the bees' death in their publications?
Then the salient paradoxical dynamics has been highlighted with the
causal factors explaining the abnormal collapse of the bees. Usually
opinion A and B are axiomatically asserted. But the empirical results
have shown that if we consider opinion A "bees are dying because of
chemicals", then opinion B is not "not A". Indeed, the dynamics described
by the "no proof" assertion can hardly explain all the shift in the
opinion dynamics of the whole debate. In reality, opinion B appears to be
"the multifactor explanation of the bees' death".

Second result has heuristic consequences. With the modelling, we have
discovered that most of the "lobbying" effort may have been performed
by the multifactor side and not by the unifactor one (mainly during
the 1998-2004 years). Authors of papers have been selected as relevant
entities to reproduce the phenomenon with a dynamical model as a
possible explanation of the dynamics of the collective judgment: the
Galam sequential probabilistic model of opinion dynamics. According to
this model, proportions of inflexible and flexible agents describe the
shifts observed in the dynamics opinion. The counter-intuitive results
show the necessary minimal proportions to maintain or to break the
dynamics. Finally, the key result of this study stresses that without
supposing any social meta-category, the proportions of multi/uni-factor
categorized papers are not the same as multi/uni-factor-inflexible
agents.

Third result has sociological implications. If the debate starts
with a rather neutral position of opinion, the debate seems to get
a strong coloration with the first assertions. Results show that
inside one phase (1998-2004 or 2005-2010), it appears very difficult
to fight against inertia of the no falsification process because of
the evolution of the thematization, as text mining analysis already
suggested~\cite{Delanoe:2010}: it becomes too "expansive" the fact of
falsificating it. This fact depends on the problem formulation which
oriented the controversy in one exclusive way~\cite{Gusfield:1981:eng}.
Here is the main difference between two definitions of the public
problem: "Do pesticides kill the bees?" or "what do kill the bees?".
The first assertion of the whistleblowers oriented the debate focusing
on the pesticides, reversing the burden of the proof. Moreover the
evolution of the framing argumentation during the first period of
the debate has implied its no-falsifiability from the scientific and
analytic field to the moral and synthetic judgement. This paper finally
concludes the agents not only select or calculate, but prioritize and
above all translate the causal factors explaining the issue: "simple
agents" become "symbolic actors" in a drama.

Fourth result deals with the parameters of the model itself. One may
postulate the data could be reproduced with groups of size 4 (or more)
since journalists can meet (or study) more than 3 opinions. Besides,
number of iterations (5 in this paper) can be modified according to the
tendencies observed with the text mining methodology. This parameters
will be changed preserving the dynamics observed. But, in fact, we found
the minimal conditions to reproduce the states. However a change in the
proportions of inflexible agents can create redundancies and heavy
investments. Finally, such modeling completed with an ethnographic and
sociological investigation could focus on the shape of the networks of
inflexible agents from both side and find how it is correlated to the
structure of debates' dynamics.

\bibliographystyle{unsrt}
\bibliography{bibliography}

\begin{thebibliography}{10}

\bibitem{Callon:1986}
M.~Callon.
\newblock {\em J. Law, Power, action and belief: a new sociology of
  knowledge?}, chapter Some elements of a sociology of translation:
  domestication of the scallops and the fishermen of St Brieuc Bay, pages
  196--223.
\newblock Routledge, London, 1986.

\bibitem{Chateauraynaud:2004}
F.~Chateauraynaud.
\newblock {\em La croyance et l'enqu\^ete}, volume~XV, chapter L'\'epreuve du
  tangible. Exp\'eriences de l'enqu\^ete et surgissements de la preuve.
\newblock Raisons pratiques, Paris, 2004.

\bibitem{Maxim:Sluijs:2010}
L.~Maxim and J.~P. van~der Sluijs.
\newblock Expert explanations of honeybee losses in areas of extensive
  agriculture in france: Gaucho ® compared with other supposed causal factors.
\newblock {\em Environmental Research Letters}, 5(1):014006, 2010.

\bibitem{Delanoe:2004}
A.~Delano\"e.
\newblock Quand les abeilles meurent les articles sont compt\'es,
  g\'en\'ealogie et analyse s\'emantique d'une crise m\'ediatique.
\newblock {\em Strategic, Scientific and Technological Watch}, 2004.

\bibitem{Delanoe:2007}
A.~Delano\"e.
\newblock Analyse des dynamiques manag\'eriales face à la contestation sociale
  : \'el\'ements statistiques du cas {"ni Gaucho, ni R\'egent"}.
\newblock {\em Strategic, Scientific and Technological Watch}, 2007.

\bibitem{dewey:1927}
J.~Dewey.
\newblock {\em The public and its problems}.
\newblock Holt, New York, 1927.

\bibitem{Gusfield:1981:eng}
J.-R. Gusfield.
\newblock {\em Drinking-driving and the symbolic order. {T}he culture of public
  problems}.
\newblock The {U}niversity of {C}hicago {P}ress, 1981.

\bibitem{afssa:2008}
J.~Chiron and A.-M. Hattenberger.
\newblock Mortalit\'es, effondrements et affaiblissements des colonies
  d'abeilles.
\newblock Technical report, AFSSA, Novembre 2008.

\bibitem{Boudon:2007}
R.~Boudon.
\newblock {\em Essais sur une rationalit\'e g\'en\'erale}.
\newblock {PUF}, Paris, 2007.

\bibitem{Galam:2002}
S.~Galam.
\newblock Minority opinion spreading in random geometry.
\newblock {\em {Eur. Phys. J. B}}, (25):403--406, 2002.
\newblock Rapid Note.

\bibitem{Galam:2005}
S.~Galam.
\newblock Heterogeneous beliefs, segregation, and extremism in the making of
  public opinions.
\newblock {\em {Phys. Rev.}}, 71(046123-1-5), 2005.

\bibitem{Galam:2007}
S.~Galam and F.~Jacobs.
\newblock The role of inflexible minorities in the breaking of democratic
  opinion dynamics.
\newblock {\em {Physica A}}, (381):366--376, 2007.

\bibitem{Galam:2010}
S.~Galam.
\newblock Public debates driven by incomplete scientific data: The cases of
  evolution theory, global warming and h1n1 pandemic influenza.
\newblock {\em {Physica A}}, (389):3619­3631, 2010.

\bibitem{Radnitzky:1987:en}
G.~Radnitzky.
\newblock The "economic" approach to the philosophy of science.
\newblock {\em The British Journal for the Philosophy of Science},
  38(2):159--179, June 1987.

\bibitem{Radnitzky:1986}
G.~Radnitzky.
\newblock {\em Cost-Benefit Thinking Applied to the Methodology of Research}.
\newblock 1986.

\bibitem{Castellano:2009}
C.~Castellano, S.~Fortunato, and C.~Vittorio~Loreto.
\newblock Statistical physics of social dynamic.
\newblock {\em {Reviews of Modern Physics}}, (81):591--646, 2009.

\bibitem{Galam:2008}
S.~Galam.
\newblock Sociophysics: a review of galam models.
\newblock {\em International Journal of Modern Physics}, (C 19):409--440, 2008.

\bibitem{Delanoe:2010}
A.~Delano\"e.
\newblock Statistique textuelle et s\'eries chronologiques sur un corpus de
  presse \'ecrite. {L}e cas de la mise en application du principe de
  pr\'ecaution.
\newblock {\em JADT, Journ\'ees internationales d'Analyses statistiques des
  donn\'ees Textuelles}, 2010.

\end{thebibliography}

\end{document}